\documentclass[%
reprint,
superscriptaddress,
 amsmath,amssymb,
 aps,
]{revtex4-2}
\usepackage{graphicx}
\usepackage{dcolumn}
\usepackage{bm}
\usepackage[lofdepth,lotdepth]{subfig}	

\begin{document}

\preprint{APS/123-QED}

\title{The onset of turbulence in particle-laden pipe flows}

\author{Willian Hogendoorn}
\affiliation{
Delft University of Technology, Multiphase Systems (3ME-P\&E), Leeghwaterstraat 39, 2628 CB Delft, The Netherlands\\
}

\author{Bidhan Chandra}
\affiliation{
Delft University of Technology, Multiphase Systems (3ME-P\&E), Leeghwaterstraat 39, 2628 CB Delft, The Netherlands\\
}
\affiliation{Indian Institute of Science Education \& Research, Chemical Engineering, Bhopal 462066, India}

\author{Christian Poelma}
\email{c.poelma@tudelft.nl}
\affiliation{
Delft University of Technology, Multiphase Systems (3ME-P\&E), Leeghwaterstraat 39, 2628 CB Delft, The Netherlands\\
}

\date{\today}

\begin{abstract}
We propose a scaling law for the onset of turbulence in pipe flow of neutrally buoyant suspensions. This scaling law, based on a large set of experimental data, relates the amplitude of the particle-induced perturbations ($\epsilon$) to the critical suspension Reynolds number, $Re_{s,c}$. Here $\epsilon$ is a function of the particle-to-pipe diameter ratio and the volume fraction of the suspended particles, $\epsilon = (d/D)^{1/2} \phi^{1/6}$. $Re_{s,c}$ is found to scale as $\epsilon^{-1}$. Furthermore, the perturbation amplitude allows a distinction between classical, intermediate and particle-induced transition.
\end{abstract}

\maketitle


The recurrent question when and how pipe flow transitions to turbulence long predates the seminal pipe flow experiments by Osborne Reynolds in the early 1880s \cite{reynolds1883xxix}.
Since pipe flow is linearly stable, finite amplitude perturbations are thus responsible for this onset of turbulence \cite{drazin2004hydrodynamic, kerswell2005recent}.
For increasing Reynolds number ($Re = UD/\nu$; with $U$ the bulk velocity, $D$ the pipe diameter and $\nu$ the kinematic viscosity), smaller perturbation amplitudes are sufficient to initiate this transition \cite{hof2003scaling}.
Initially, turbulence is found to be transient: localized patches of turbulence are embedded in a laminar flow.
These turbulent puffs are known to have an increasing lifetime for increasing Reynolds number \cite{kuik2010quantitative, hof2006finite}.
Beyond a Reynolds number of 2040, puffs grow and split, eventually leading to sustained turbulence \cite{avila2011onset}.\\

Particle-laden flows are known to exhibit significantly different transition behavior \cite{matas2003transition, hogendoorn2018particle, agrawal2019transition, leskovec2020pipe}.
This is in particular evident from the critical Reynolds number, $Re_c$, which is known to strongly depend on the particle volume fraction ($\phi$) and the particle-to-pipe diameter ratio ($d/D$) of the suspended particles.
For increasing volume fraction and a large enough $d/D$, critical (suspension) Reynolds numbers as low as 600 are reported \cite{matas2003transition}.
Furthermore, beyond a critical volume fraction, which depends on the particle-to-pipe diameter ratio, a different regime is observed: the transition is smooth without the presence of turbulent puffs \cite{hogendoorn2018particle, agrawal2019transition}.
The exact onset of turbulence is of major importance for a variety of applications, as this onset is accompanied by a significant - sometimes intermittent - drag increase.
Precise control over the flow rate (by means of a driving pressure) is essential for process control, whether it is in additive manufacturing, food processing, slurry transport, or dredging.
However, despite considerable efforts, a definite scaling indicating the onset of the drag increase in the transitional regime is still absent.
We propose a scaling for $Re_c$ in neutrally buoyant suspensions, which is supported by a large set of experimental data.\\

The first detailed study reporting a prominent effect of particles on laminar-turbulent transition was performed by \citet{matas2003transition}.
They determined $Re_c$ for a wide range of $d/D$ and $\phi$, using the low frequency component in the pressure spectrum as an indicator for the presence of turbulent puffs, characteristic for the onset of turbulence.
This $Re_c$ was based on the corrected viscosity using Kriegers' viscosity model \cite{krieger1972rheology}, to account for the presence of particles.
A scaling in terms of $\phi D/d$ as function of this viscosity corrected critical Reynolds number ($Re_{s,c}$) was proposed to collapse all results on a single master curve.
However, the ratio $D/d$ is used in both axes of this master curve, suggesting that $Re_{s,c}$ is a function of $\phi$ only.

\citet{lashgari2014laminar} studied the influence of neutrally buoyant particles ($d/h$ = 0.1, with $h$ the channel height) numerically for a channel flow configuration.
They introduced a distinction based on the dominant term in the stress budget.
Three different regimes are identified: a laminar regime for low volume fractions and low Reynolds numbers, a turbulent regime for low volume fractions and high Reynolds numbers, and an inertial shear-thickening regime for $\phi \gtrapprox$ 0.15.
This distinction is only feasible in numerical studies as it requires very detailed flow information.

A different transition mechanism, without the presence of turbulent puffs in the transition region, was found by \citet{hogendoorn2018particle} for higher volume fractions ($\phi \geq$ 0.175).
This particle-induced transition behavior is characterised by a smooth transition curve, which collapses on $64/Re$ for low $Re_s$ after viscosity correction.
The onset of turbulence was identified using a 10\% deviation from the law of Hagen-Poiseuille.
\citet{agrawal2019transition} independently reported this particle-induced transition for higher volume fractions as well.

This smooth, particle-induced transition was also found for lower volume fractions ($\phi$ = 0.05) in combination with a larger particle-to-pipe diameter ratio ($d/D$ = 0.17) by \citet{leskovec2020pipe}.
The authors proposed a scaling to distinguish between classical and particle-induced transition based on a model of viscous dissipation and particle agitation.

\citet{hogendoorn2021suspension} showed that for large $d/D$ even very dilute systems ($\phi$ = 0.0025) exhibit this particle-induced transition.
Based on instantaneous velocity measurements they showed that particles introduce perturbations and formulated a model predicting that these disturbances are proportional to $d/D$ and $U$.

Based on these previous studies, it appears that the onset of turbulence in particle-laden flows is dependent on at least two parameters: the particle-to-pipe diameter ratio and the volume fraction.
We propose a scaling for this onset in neutrally buoyant suspensions, based on a wide range of experimental data, both new and from aforementioned studies.
Furthermore, this scaling can be used to predict which transition behavior will be observed for a given system.\\

Experiments are performed in two different pipe-flow facilities.
The first experimental setup is the same as the setup described in \citet{hogendoorn2018particle}.
In short, this setup consist of a precision glass pipe with a diameter ($D$) of 10.00$\pm$0.01 mm.
The flow is gravity driven using an overflow reservoir to prevent perturbations from the pumps.
The height of this reservoir can be adjusted to control $Re_s$.
At the inlet a settling chamber and a smooth contraction are used to maintain (single-phase) laminar flows for Reynolds numbers exceeding 4000.
An orifice, similar to the one used by \citet{wygnanski1973transition}, is used to ensure a controlled transition at a fixed Reynolds number of 2000 for single-phase flows.
The total pipe length ($L$) after the orifice is 310$D$.
The pressure drop ($\Delta p$) is measured (Validyne DP15) from 125$D$ to 250$D$, ensuring sufficient development length.
$Re_s$ is determined with an uncertainty smaller than 0.5\%, by collecting and weighing an amount of suspension from the outflow during a given time.
A set of peristaltic pumps are used to feed the outflow back to the overflow reservoir.

The second experimental setup is similar to the one described above; for brevity the differences will be addressed only.
This setup consist of a PMMA pipe with an inner diameter of 19.98$\pm$0.06 mm.
Using a settling chamber with a smooth contraction in combination with smooth pipe connectors a laminar flow is maintained for single-phase Reynolds numbers up to 5000.
In this setup the flow is either perturbed using an orifice or perturbed using an active perturbation mechanism, which is a zero-net mass flux injector (adapted from \citet{draad1998laminar}) at the beginning of the pipe (positioned 10$D$ after the inlet chamber).
The perturbation method is no longer significant beyond a certain critical volume fraction as was reported by \citet{matas2003transition} and \citet{agrawal2019transition}.
We also confirmed this for our experiments.
The average pressure drop is obtained from 125$D$ to 200$D$ after this active perturbation.
An inline Coriolis mass flow meter (KROHNE Optimass 7050c) is used to measure the flow rate with a maximum uncertainty of $\pm$ 0.1\%.
A progressive cavity pump (Monopump, AxFlow B.V.) is used to transport the suspension back to the feeding reservoir.
For $d/D$ = 0.088, the overflow was removed to be able to drive very viscous flows. We confirmed that this did not influence the (single-phase) transition, as this was still dominated by the orifice perturbation.

Saline water (Na\textsubscript{2}SO\textsubscript{4}) or a water-glycerine mixture is used to obtain a density-matched system with polystyrene particles (Synthos EPS; density $\rho$ = 1.032 kg/L).
The volume fraction in the experimental facilities is controlled based on the mass ratio of the working fluid and the particles. 
Starting with a single-phase system, with a known initial mass, particles are added in steps to obtain the desired volume fraction.
Particles with diameters of 0.30$\pm$0.034, 0.53$\pm$0.05, 1.31$\pm$0.07 and 1.75$\pm$0.12 mm are used.
All four particle types are used in the 10.00 mm facility and the 1.31 and 1.75 mm particles are also used in the 19.98 mm facility.
This results in six $d/D$ ratios (0.03, 0.053, 0.065, 0.088, 0.13, and 0.18); two pressure data sets were re-used from previous studies ($d/D$ = 0.18, 0.053 \cite{hogendoorn2018particle, hogendoorn2021suspension}).

The (critical) Reynolds numbers reported in this study are based on the corrected viscosity ($\mu_s = \rho\nu_s$) of the suspension, determined using Eilers' viscosity model \cite{stickel2005fluid}:
\begin{equation}
\frac{\mu_s}{\mu_0} = \bigg(1 + 1.25\frac{\phi}{1-\phi/0.64}\bigg)^2.
\label{eq:1}
\end{equation}
Here $\mu_0$ is the viscosity of the continuous phase (i.e., saline water or glycerol).

\begin{figure}[htb]
\includegraphics[width=\linewidth]{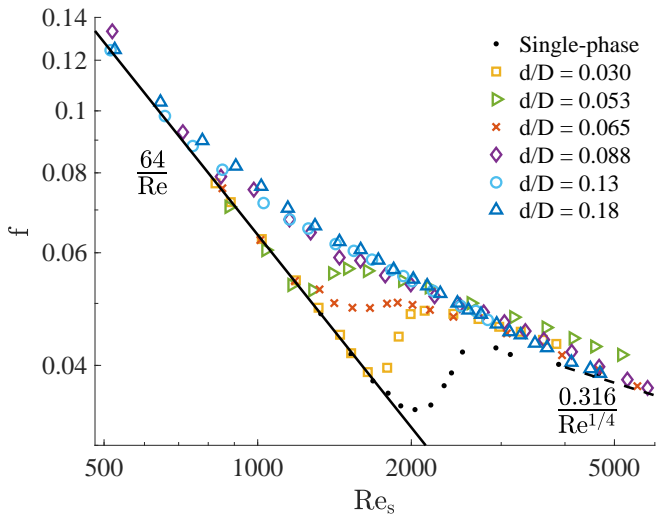}
\caption{\label{fig:fig1} Friction factor, $f$, as function of suspension Reynolds number, $Re_s$. The single-phase case is shown for reference. For the particle-laden cases the concentration ($\phi$) is fixed at 0.05 to highlight the diameter-ratio ($d/D$) effect.}
\end{figure}

Fig.~\ref{fig:fig1} shows the transition behavior for a range of different experiments (i.e., various $d/D$) for a constant volume fraction, $\phi$ = 0.05.
Here the Darcy friction factor ($f \equiv \Delta p/(\frac{1}{2}\rho U^2 L/D)$) is shown as function of $Re_s$.
The continuous line represents the Hagen-Poiseuille law: $64/Re$, the solution for laminar flows. The dashed line shows Blasius' equation.
The single-phase case obtained in the 10.00 mm diameter setup, shown for reference, displays a transition at $Re_{c,0} \approx$ 2000, resulting from the perturbation in the beginning of the setup.
For the particle-laden cases, the influence of $d/D$ is clearly visible from the decrease of $Re_{s,c}$ for increasing $d/D$.
For $d/D <$ 0.065, a sharp transition is observed with a clear local minimum in the transition region.
This local minimum shifts to lower $Re_{s,c}$ for increasing $d/D$, implying an earlier onset of turbulence.
A critical transition curve is shown for $d/D$ = 0.065, where the local minimum is still (only just) present.
Eventually, for $d/D >$ 0.065, smooth transition curves are observed, characteristic for particle-induced transition.

Notably, for the intermediate case ($d/D$ = 0.065), the friction factors in the transition region are {\em{lower}} compared to the friction factors corresponding to $d/D$ = 0.053.
Related to this change in transition behavior is the non-monotonically decreasing critical Reynolds number in this specific region.
This can likely be explained by the change in transition scenario and the associated change in length scales.
In a previous study, the integral length scales corresponding to a smooth transition case were found to be smaller (i.e., approximately 4$D$ in the transition region) and continuously present \cite{hogendoorn2021suspension}.
This is in contrast to a sharp, intermittent transition, where turbulent patches span about 20-30$D$ \cite{wygnanski1973transition, eckhardt2007turbulence}.

\begin{figure}[htb]
\includegraphics[width=\linewidth]{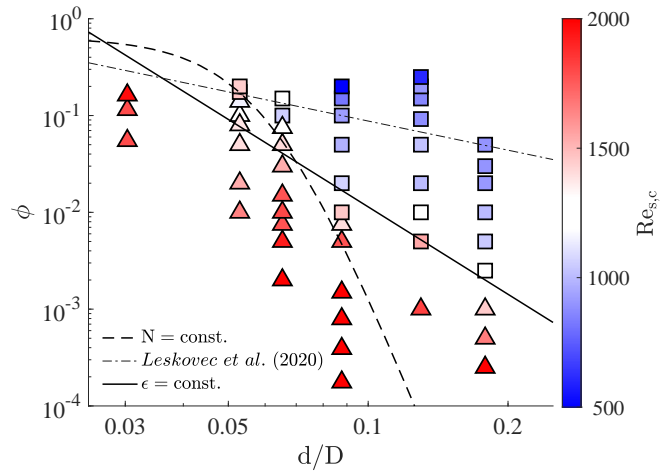}
\caption{\label{fig:fig2} Regime map $\phi$ vs $d/D$, where each marker represents one complete transition curve.
The transition curves indicated with the triangular markers have a local minimum, whereas the transition curves with a monotonically decreasing friction factor are represented by a square marker.
The marker color represents the critical suspension Reynolds number, $Re_{s,c}$.}
\end{figure}

All data are shown in the regime map ($\phi$ vs.~$d/D$) in Fig.~\ref{fig:fig2}, where each marker represents one of 51 complete transition curves (which consist of 11-39 measurements of $f$ vs.~$Re_s$).
The transition curves with a monotonically decreasing friction factor for increasing $Re_s$ (i.e., $\partial f/\partial Re < 0$ for all values of  $Re$) are indicated with square markers.
The transition curves with a local minimum are represented by the triangular markers.
Here the derivative is locally positive in the transition region.
The color of the markers indicates $Re_{s,c}$, determined using a threshold of $70/Re$ \cite{hogendoorn2018particle}.
This threshold was determined to be a sound balance between robustness to measurement noise and accuracy.
This figure confirms the dependency of $Re_{s,c}$ on $d/D$ and $\phi$: an increase in either parameter promotes transition. 
The dashed curve represents a constant Bagnold number \cite{bagnold1954experiments}, which has previously been used to classify suspension behavior \cite{lashgari2014laminar}.
It is defined as the ratio of the inertial to viscous stress: $N = d^2 \dot{\gamma} \lambda^{1/2}/\nu$, with $\dot{\gamma}$ the shear-rate and $\lambda$ the linear concentration, $1/[(0.74/\phi)^{1/3}-1]$.
The best discrimination between transition mechanisms is found for $N$ = 7.2 (based on a bulk shear-rate for $Re_{c,0}$ = 2000).
It is evident that this is still not satisfactory.
Alternative values of $N$ will always only satisfy the transition behavior at one $d/D$.
We thus confirm the observation by \citet{lashgari2014laminar} that the Bagnold number by itself is not sufficient to predict transition behavior.
All experiments shown in Fig.~\ref{fig:fig2} are well below $N$ = 40, which suggest that all cases are in the viscous-dominated regime according to Bagnolds' theory.
Another model, based on particle agitation versus laminar-dissipation is proposed by \citet{leskovec2020pipe}.
According to their model the threshold between the two mechanisms is predicted by $\phi^2(d/D)^2Re$ and is indicated by the dashed-dotted curve.
However, this model is based on a limited range of experimental data, resulting in a less accurate prediction for the transition mechanism at higher $d/D$.
The solid curve, indicated by $\epsilon$ = const., is based on our proposed model (introduced below) to distinguish between intermediate and particle-induced transition.\\

\begin{figure*}[t]
    \centering
    \subfloat[]{
    \includegraphics[width=0.47\textwidth]{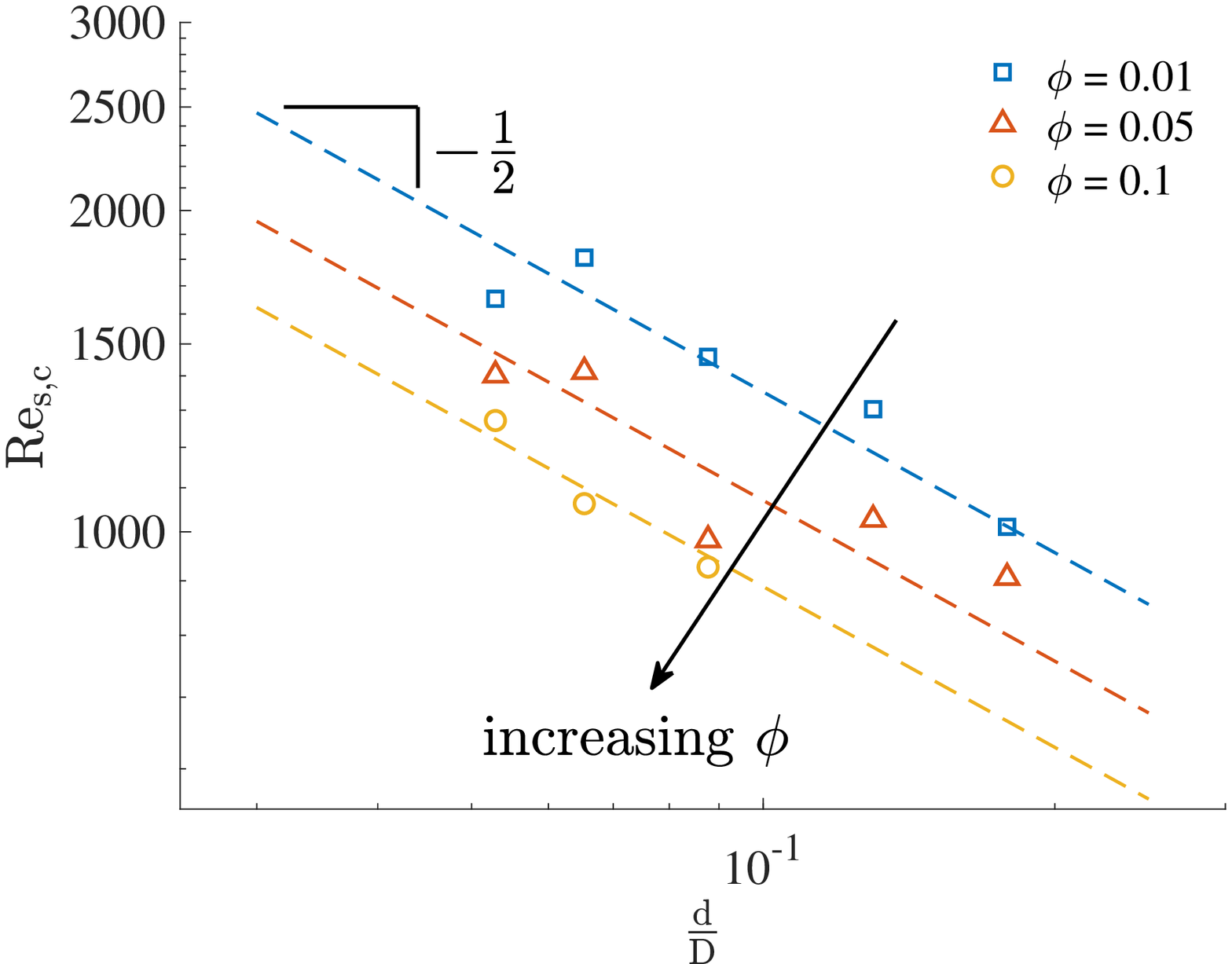}
    \label{fig:3a}}
    \subfloat[]{
    \includegraphics[width=0.47\textwidth]{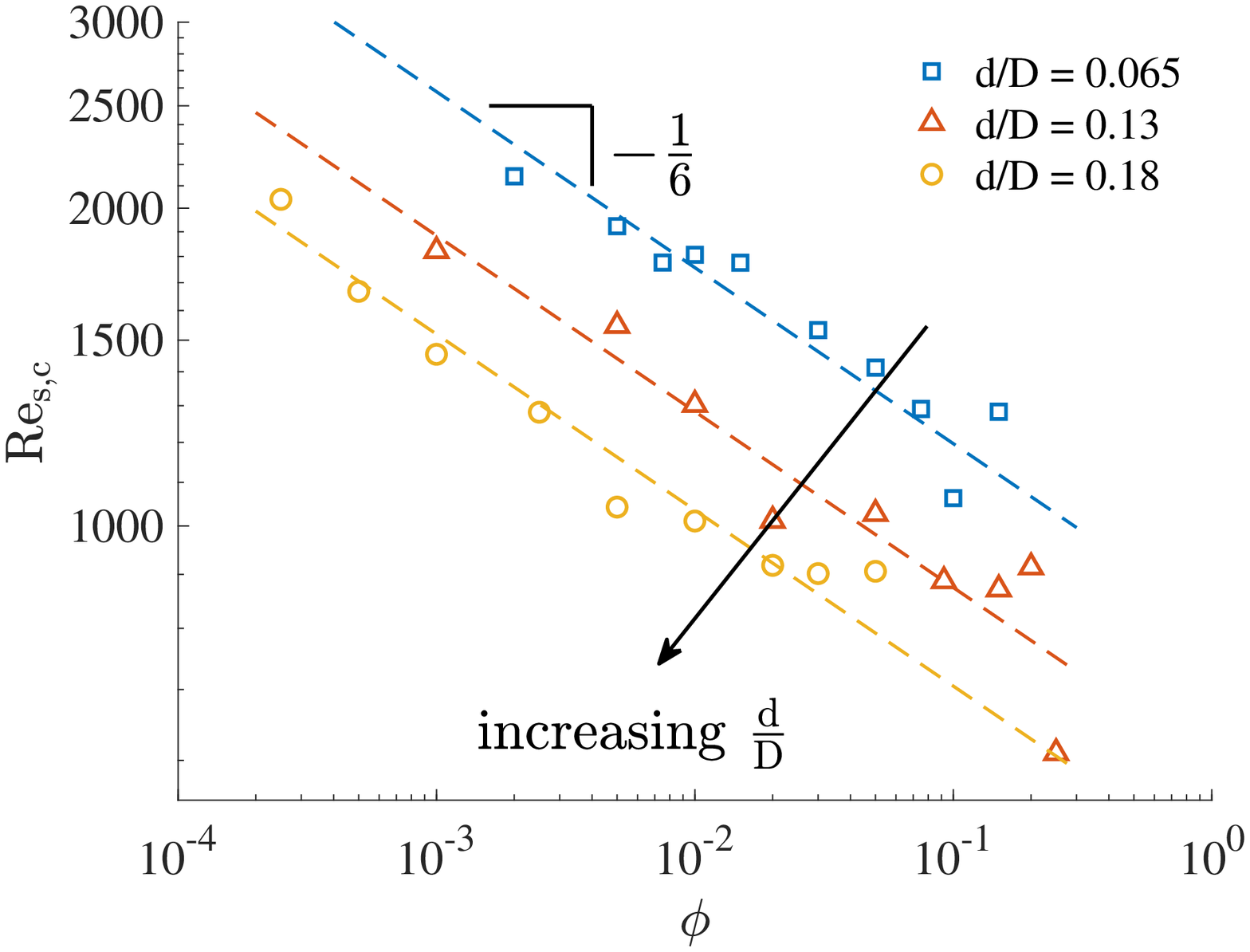}
    \label{fig:3b}}
    \caption{\protect\subref{fig:3a} $Re_{s,c}$ as function of $d/D$ for three representative volume fractions and \protect\subref{fig:3b} $Re_{s,c}$ as function of $\phi$ for three representative diameter ratios. The dashed curves are based on a regression analysis, where the slope is rounded to the nearest common fraction.}
    \label{fig:3}
\end{figure*}

Previously, it has been shown that the velocity fluctuations caused by the finite-sized particles scale with $d/D$.
Based on this, it is expected that $Re_{s,c}$ is dependent on $d/D$ and $\phi$ \cite{hogendoorn2021suspension}.
The latter has been shown to be a relevant scaling parameter before \cite{matas2003transition, hogendoorn2018particle, agrawal2019transition}.
The exact respective contributions of $d/D$ and $\phi$ on $Re_{s,c}$ are studied, based on the data  shown in Fig.~\ref{fig:fig2}.
The effect of $d/D$ (for $\phi$ = 0.01, 0.05, and 0.1) and $\phi$ (for $d/D$ = 0.065, 0.13, and 0.18) are shown in Fig.~\ref{fig:3a} and \ref{fig:3b}, respectively. In both figures a selection of three representative cases is shown.
From both figures it follows that $Re_{s,c}$ depends on $d/D$ and $\phi$ with exponents of $-\frac{1}{2}$ and $-\frac{1}{6}$, respectively.
The exponents are determined using a regression analysis, where the found exponents are rounded to the nearest common fraction.
The dashed curves shown are based on this fraction and the corresponding amplitude, resulting from the regression analysis. While both exponents are obtained separately in this approach, we also confirmed that a regression of both exponents simultaneously, ($d/D)^\alpha\phi^\beta$, gave similar results. Note that the square of the linear concentration in Bagnolds' scaling also results in an exponent of $\frac{1}{6}$ for the concentration in the dilute limit.

Based on this, the onset of turbulence is modelled as $Re_{s,c} \big(\frac{d}{D}\big)^\frac{1}{2} \phi^\frac{1}{6} = c$.
Note that dimensionless group, $\big(\frac{d}{D}\big)^\frac{1}{2} \phi^\frac{1}{6}$, can also be interpreted as a perturbation amplitude, $\epsilon$, resulting from the suspended particles.
This is analogous to finite perturbation amplitudes being used in single-phase transition experiments \cite{hof2003scaling, waleffe1997self, mullin2011experimental}. 
Therefore we decided to introduce the following perturbation amplitude:
\begin{equation}
\epsilon = \Big(\frac{d}{D}\Big)^{\frac{1}{2}} \phi^{\frac{1}{6}}.
\label{eq:2}
\end{equation}
This perturbation amplitude can also be rewritten to be proportional to $N_p^{~\frac{1}{6}} \frac{d}{D}$, with $N_p$ the number of particles per unit volume (i.e., $D^3$).
We choose the former representation to separate the parameters, as $N_p$ is a function of both $d/D$ and $\phi$.
The physical interpretation of $\epsilon$ is that the perturbation amplitude increases with the number of particles per unit volume and for increasing $d/D$.

\begin{figure}[htb]
\includegraphics[width=\linewidth]{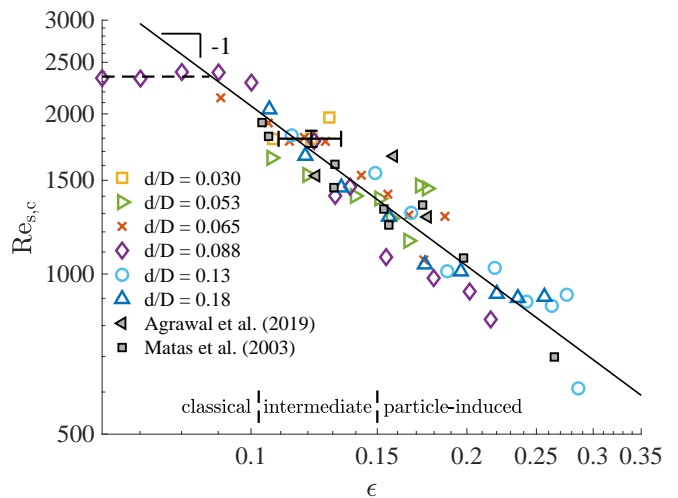}
\caption{\label{fig:fig4} Critical suspension Reynolds number, $Re_{s,c}$, as function of particle perturbation amplitude, $\epsilon$, for all experiments. All $Re_{s,c}$ collapse on one single curve with slope -1. The marker on the ordinate axis represents the critical Reynolds number for a single-phase flow, $Re_{c,0}$.}
\end{figure}

In Fig.~\ref{fig:fig4}, all critical Reynolds numbers (i.e., the colors from Fig.~\ref{fig:fig2}) are shown as function of $\epsilon$.
As expected, for this scaling all $Re_{s,c}$ now collapse on one single curve given by: $Re_{s,c} = 207~\epsilon^{-1}$.
The general interpretation of the exponent in case of single-phase flows is that there is a balance between inertial and viscous forces \cite{mullin2011experimental}. For the particle-laden cases the exponents for $d/D$ and $\phi$ are now responsible for this balance.
The pre-factor, also resulting from regression, is likely specific for the current configuration: the flow of a suspension of neutrally buoyant, spherical particles through a pipe. 
The relatively large horizontal errorbar, shown for one experiment only, is based on a conservative error propagation and predominantly originates from the polydispersity in particle diameter (common for experimental studies).
This new scaling is also validated using data from literature \cite{matas2003transition,agrawal2019transition}. These $Re_{s,c}$ are indicated in the legend.
For $Re_{s,c}$ reproduced from \citet{matas2003transition}, three $Re_{s,c}$ for three different $d/D$ (i.e., 0.056, 0.063 and 0.1) are shown, spanning a significant range of the scaling.
Note that $Re_{s,c}$ beyond the local minimum (see Fig.~3 in \citet{matas2003transition}) are excluded from this analysis as these $Re_{s,c}$ are biased due to their measurement method, as is discussed in \citet{hogendoorn2018particle}.
Data for $Re_{s,c}$ taken from \citet{agrawal2019transition} also support our scaling.

For small $\epsilon$ (i.e., low $\phi$ and/or $d/D$), particles will not affect $Re_{s,c}$.
In this regime, the amplitude of the particle perturbations is negligible with respect to the perturbation amplitude of the disturbance mechanism.
Therefore, the transition behavior will be described by the dashed line in Fig.~\ref{fig:fig4}, indicating a fixed transition corresponding to the perturbation amplitude of the used perturbation mechanisms ($Re_{c,0}$ = 2350 for this particular experiment).
The presence of this plateau is confirmed for experiments with a particle size ratio of $d/D$ = 0.088 (see also $Re_{s,c}$ for values of $\epsilon \rightarrow$ 0 in Fig.~3 in \citet{matas2003transition}).
The marker on the ordinate axis represents the corresponding critical Reynolds number for this single-phase case ($Re_{c,0}$), i.e., $\epsilon \rightarrow$ 0.

Using the scaling law, the conditions can be identified where the perturbations of the particles are sufficiently damped by viscous effects, so that transition is not triggered by the particles.
For combinations of $\epsilon$ and $Re_s$ above the solid line in Fig.~\ref{fig:fig4}, the friction factor will deviate 10\% or more from Poiseuille. Conversely, for combinations below this line, the friction factor can safely be assumed to be $64/Re_s$.

The scaling shown in Fig.~\ref{fig:fig4} covers a very wide range of situations, including most engineering applications. As an indication, a 45\% volume fraction of particles with a size of 1/10 of the pipe diameter is characterized by $\epsilon$ = 0.28. Larger values can be reached by increasing $d/D$ to unity (leading to a theoretical limit of $\epsilon$ = 0.93). However, a different scaling is then expected, as these extreme cases would represent stacked particles with a diameter close to the pipe diameter. Obviously, this is no longer a `flowing suspension'.

In the literature generally three different regimes are distinguished to describe the underlying dynamics: classical, intermediate, and particle-induced transition behavior \cite{hogendoorn2018particle, agrawal2019transition, hogendoorn2021suspension}.
Using the perturbation amplitude (Eq.~\ref{eq:2}), we can quantitatively distinguish between these different regimes; they are indicated in the bottom of Fig.~\ref{fig:fig4}.
The border between classical and intermediate transition can be defined where $Re_c$ is found to deviate from a typical transition Reynolds number for single-phase flow, $Re_{c,0}$ = 2000.
The corresponding critical perturbation amplitude is found to be $\epsilon$ = 0.103.
The change between intermediate and particle-induced transition behavior is indicated by the solid curve in Fig.~\ref{fig:fig2}, corresponding to $\epsilon$ = 0.15.
This value is determined by minimizing the error between the number of square markers and triangular markers above and below the curve, respectively.
Given the limited number of data near the prediction curve, this critical value for $\epsilon$ is bounded by 0.135 and 0.17.
For simplicity we report a value in the center of this range.
Note that there is a smooth transition between classical and particle-induced transition behavior for increasing $\epsilon$.
The intermittent nature of classical transition is gradually replaced by continuous, particle-induced fluctuations, see also the detailed characterization of one single case by \citet{hogendoorn2021suspension}.

Our motivation to interpret $\epsilon$ as a perturbation amplitude follows from similar approaches in single-phase flow experiments \cite{waleffe1997self, mullin2011experimental}.
Note that for suspension flows the particle-induced perturbations are continuously present along the length of the pipe.
This is in contrast to single-phase perturbation experiments, where the perturbation is temporally and spatially bounded.
For injection-based disturbances the perturbation amplitude is typically defined as the ratio of the injection volume flux with respect to the volume flux in the pipe.
Similarly, for an orifice type perturbation the orifice diameter can be expressed as a disturbance amplitude.
The amplitude required to trigger transition in single-phase flows scales with $Re^{\gamma}$, where the exponent varies between -1 and -1.5, depending on the type of perturbation \cite{hof2003scaling, peixinho2007finite, mullin2011experimental}.
This scaling is generally based on experiments for $Re >$ 2000, while for particle-laden flows the range {\em{below}} $Re_s$ = 2000 is especially important.

Further measurements with small $d/D$ and/or low $\phi$ (i.e., low $\epsilon$) in the absence of a perturbation mechanism should reveal the flow stability response for $Re_{s,c} >$ 2000.
This will also shed light on the universality of the various perturbation parameters.
On the edges of the investigated parameter space (Fig.~\ref{fig:fig2}) various effects will come into play \cite{morris2020toward}. For instance, the effect of the spatial distribution of particles on the transition in the (semi-)dilute regime is still an open question.
For $d/D \rightarrow$ 0, our model predicts that particles will not affect the transition, but this needs to be confirmed.
Additionally, measurements for higher volume fractions need to be performed to establish whether our scaling law will hold in the inertial-shear thickening regime \cite{lashgari2014laminar}.\\

In summary, based on a large set of experimental data, we uncovered a scaling law relating the amplitude of the particle-induced perturbations to the critical suspension Reynolds number.
The particle-induced perturbation amplitude is a simple function of the particle-to-pipe diameter ratio and the volume fraction: $\epsilon = (d/D)^{1/2} \phi^{1/6}$.
The onset of turbulence in neutrally buoyant suspensions is found to scale as $Re_{s,c} \sim \epsilon^{-1}$.
Data from literature also supports the validity of this scaling.
Furthermore, $\epsilon$ allows a prediction of the transition scenario.
For a variety of applications it will predict whether the transition will be classical, intermediate, or particle-induced.\\

\begin{acknowledgments}
We thank Vasudevan Krishnan for assisting in our experimental campaign. We would like to thank Mr S. van Baal (Synthos EPS) for kindly providing particles for this research. This work is funded by the ERC Consolidator Grant No. 725183 “OpaqueFlows.”
Underlying data are deposited in the 4TU.Centre for Research Data, and will be accessible by: \href{http://doi.org/10.4121/16586954}{doi:10.4121/16586954}.\end{acknowledgments}


\bibliography{bibliography}

\end{document}